\documentclass[pdflatex,sn-mathphys-num]{sn-jnl}

\usepackage{graphicx}
\usepackage{multirow}
\usepackage{amsmath,amssymb,amsfonts}
\usepackage{amsthm}
\usepackage{mathrsfs}
\usepackage[title]{appendix}
\usepackage{xcolor}
\usepackage{textcomp}
\usepackage{manyfoot}
\usepackage{booktabs}
\usepackage{algorithm}
\usepackage{algorithmicx}
\usepackage{algpseudocode}
\usepackage{listings}
\usepackage{pdflscape}
\usepackage{caption}
\usepackage{rotating}
\usepackage{siunitx}
\usepackage[version=4]{mhchem}
\usepackage{graphicx} 
\usepackage{subcaption}
\usepackage{float}
\usepackage[table]{xcolor}
\usepackage{siunitx}
\usepackage{mhchem}
\PassOptionsToPackage{hyphens}{url}\usepackage{url}
\usepackage{cleveref}
\usepackage{makecell}

\usepackage{titlesec}

\titleclass{\subsubsubsection}{straight}[\subsubsection]
\newcounter{subsubsubsection}[subsubsection]
\renewcommand\thesubsubsubsection{\thesubsubsection.\arabic{subsubsubsection}}
\titleformat{\subsubsubsection}
  {\normalfont\normalsize\bfseries}{\thesubsubsubsection}{1em}{}
\titlespacing*{\subsubsubsection}{0pt}{3.25ex plus 1ex minus .2ex}{1.5ex plus .2ex}

\begin{document}

\title{PET Attenuation Map Generation for Brain Imaging from Virtual Non-Contrast and Virtual Non-Iodine Images Obtained with Photon Counting CT Scanners}

\author*[1,2]{\fnm{Heba} \sur{Alrakh}}\email{h.alrakh@fz-juelich.de}

\author[1]{\fnm{Lutz} \sur{Tellmann}}

\author[3]{\fnm{Omid} \sur{Nikoubashman}}

\author[3]{\fnm{Sadaf} \sur{Shokrgozar}}

\author[1]{\fnm{Cláudia} \sur{Régio Brambilla}}

\author[1]{\fnm{Ahlam} \sur{Said Mohamad Issa}}

\author[1]{\fnm{Keith George} \sur{Ciantar}}

\author[4]{\fnm{Sebastian} \sur{Faby}}

\author[5]{\fnm{Ali} \sur{Abu arra}}

\author[3]{\fnm{Martin} \sur{Wiesmann}}

\author[3]{\fnm{Charlotte} \sur{Weyland}}

\author[1,2]{\fnm{N.~Jon} \sur{Shah}}

\author[1]{\fnm{Christoph} \sur{Lerche}}\email{c.lerche@fz-juelich.de}

\affil*[1]{\orgdiv{Institute of Neuroscience and Medicine (INM-4)}, \orgname{Forschungszentrum Jülich GmbH}, \orgaddress{\street{Wilhelm-Johnen-Straße}, \city{Jülich}, \postcode{52428}, \state{North Rhine-Westphalia}, \country{Germany}}}

\affil[2]{\orgdiv{Department of Neurology}, \orgname{RWTH Aachen University}, \orgaddress{\street{Pauwelsstraße 30}, \city{Aachen}, \postcode{52074}, \state{North Rhine-Westphalia}, \country{Germany}}}

\affil[3]{\orgdiv{Department of Diagnostic and Interventional Neuroradiology}, \orgname{RWTH Aachen University}, \orgaddress{\street{Pauwelsstraße 30}, \city{Aachen}, \postcode{52074}, \state{North Rhine-Westphalia}, \country{Germany}}}

\affil[4]{\orgdiv{Department of Computed Tomography}, \orgname{Siemens Healthineers AG}, \orgaddress{\street{Siemensstr. 3}, \city{Forchheim}, \postcode{91301}, \state{Bavaria}, \country{Germany}}}

\affil[5]{\orgdiv{Department of Biophysics and Medical Imaging}, \orgname{An-Najah National University}, \city{Nablus}, \country{Palestine}}

\abstract{\indent \textbf{Purpose} This study presents a transformation scheme for converting measured Hounsfield units (HUs) into linear attenuation coefficients (LACs) for \SI{511}{\keV} gamma photons in iodine contrast-enhanced angiographic CT images. The method is based on spectral CT data acquired from tissue-equivalent inserts of the Sun Nuclear electron density phantom. The resulting transformation will be used to develop an attenuation map template for our in-house developed BrainPET 7T system.

\indent \textbf{Methods} Three approaches were used to obtain the attenuation maps. The first approach is based on angiographic spectral CT scans with iodinated contrast agent administration, and the computation of virtual non-contrast (VNC) and virtual non-iodine (VNI) images. In this case, two images are combined: bone regions are segmented from the VNI images, and the corresponding voxel values in the VNC images are replaced, yielding a single quantitative monochromatic CT image. The second approach relies exclusively on VNC images, whereas the third approach uses only VNI images.

\indent \textbf{Results} The transformation schemes from HUs to LACs were established for different tube voltages (\SI{120}{\kV} and \SI{140}{\kV}) and virtual monochromatic energies (\SI{64}{\keV}, \SI{67}{\keV}, \SI{70}{\keV}, \SI{76}{\keV}, \SI{80}{\keV}, and \SI{100}{\keV}), showing the HUs dependency on the energy setting of the reconstructed virtual monochromatic energies. The accuracy and precision of the proposed transformation scheme were further evaluated using previously acquired angiographic CT datasets.

\indent \textbf{Conclusion} Spectral CT can be expected to minimize the systematic error when converting HUs to LACs for \SI{511}{\keV} gamma rays using virtual monochromatic images (VMIs). Further, using VNC images to compute PET attenuation coefficients from CT angiographic images after the administration of iodinated contrast agents allows for a simple transformation scheme, as a single linear function can be used across different voltage settings and because the LAC predictions for soft tissue are more accurate.}


\maketitle

\section{Introduction}\label{Introduction}

Quantitative positron emission tomography (PET) requires precise attenuation correction (AC), as a large fraction of the photons are scattered within the subject matter, reducing detector counts in the corresponding lines of response (LORs) and altering their apparent origin. To compensate for these effects and to enable reliable quantification, an attenuation map (AM) with the accurate \SI{511}{\keV} linear attenuation coefficients (LACs) for each voxel is required \cite{martinez-mollerAttenuationCorrectionPET2012}.

In conventional PET/CT systems, attenuation maps are typically derived from computed tomography (CT) images by converting Hounsfield Units (HUs) to LACs at \SI{511}{\keV}. Pioneering this field, Kinahan et al.~\cite{kinahanAttenuationCorrectionCombined1998} evaluated several conversion methods, including water-property-based scaling, tissue-type segmentation with assigned \SI{511}{\keV} LACs, and a hybrid approach combining both techniques. Using phantom and human data, Kinahan et al.~demonstrated that the hybrid method provides highly accurate attenuation maps. Building directly upon these findings, Carney et al.~\cite{carneyMethodTransformingCT2006} subsequently formalized the HU-to-LAC conversion that is widely used in modern clinical systems.

However, due to the fact that conventional CT uses a polyenergetic X-ray beam, while PET $\gamma$-photons are monoenergetic, these traditional conversion methods are inherently inconsistent \cite{kopsAttenuationCorrectionCerebellum2024}. Importantly, converting between two monoenergetic spectra introduces less conversion bias than converting from polyenergetic spectra to monoenergetic spectra \cite{xiaDualEnergyCT2013a}. Moreover, existing transformation schemes suffer from the difficulty that they are not simultaneously optimized for all possible kilovolt peak (kVp) settings and cannot reliably differentiate between materials, such as bone and contrast agents with similar HUs. This limitation introduces additional sources of conversion bias \cite{carneyMethodTransformingCT2006, kinahanAttenuationCorrectionCombined1998, burgerPETAttenuationCoefficients2002}. 

To address these shortcomings, advances in attenuation mapping, particularly using CT techniques, could potentially improve PET quantitation accuracy. Among these, spectral CT techniques, such as dual-energy CT (DECT) or photon-counting CT (PCCT) \cite{sodicksonDualEnergyCT2021}, enable material decomposition and other advanced applications like virtual non-contrast (VNC) and virtual non-iodine (VNI). They also facilitate the computation of virtual monoenergetic images (VMIs), providing an inherently more robust derivation of LACs from polyenergetic CT data across diverse kVp settings.  

In typical PET/CT workflows, the CT scan used for attenuation map generation is acquired shortly before the PET acquisition. In contrast, hybrid PET/MR systems, such as our in-house developed 7T BrainPET, integrate PET with magnetic resonance imaging. This combination provides superior soft tissue contrast and advanced functional imaging, without the need for ionizing radiation. However, because MR does not provide the information available from CT, the established HU-to-LAC transformation used in PET/CT cannot be directly applied to PET/MR. As a result, attenuation map generation is performed using either previously acquired CT images from the same patient, or— more commonly—segmentation, atlas-based, or template-based methods \cite{LADEFOGED2017346, wagenknechtMRIAttenuationCorrection2013, KopsTemplateBasedAttenuationCorrection, LeeReviewDeep-Learning_AC}. 

When generating attenuation atlases or templates for head PET/MR imaging with CT acquisitions, particular care must be taken regarding the eye lenses, as they are highly sensitive to ionizing radiation \cite{DoseLimit_EyeLens}. To address this concern, dedicated head CT scans are often performed with special techniques to protect the eye lenses. Common approaches include the use of bismuth eye shields or gantry tilting. While these strategies effectively reduce the CT dose, they may also affect the image quality due to additional attenuation or the restriction of the PET field of view (FOV) \cite{heaneyComparisonReductionCT2006}. Alternatively, low-dose CT is another option for generating attenuation maps; however, it may introduce quantification bias. For instance, Zhang et al.~\cite{zhangQuantitativeAccuracyCT2018} reported bias values of up to \SI{10}{\percent} for low-dose CT scans.
As a further solution, spectral CT enables the extraction of attenuation maps from iodine contrast-enhanced angiographic head images, which are acquired at a normal dose and without eye shields. By leveraging the material decomposing possibilities of spectral CT, the iodine can be suppressed, allowing for AM template generation of the head, neck, and shoulders for PET/MR imaging directly from angiographic CT images. However, in this case, Carney's transformation method cannot be applied \cite{carneyMethodTransformingCT2006} as the method employs single-kVp CT scans, and the algorithmic iodine removal for the HU-to-LAC transformation may again introduce bias. Moreover, the dependence of the transformation formula on reconstruction energy, i.e.~the virtual discrete energies of VMI, in spectral CT is expected to differ from the kVp dependence described in Carney's method.

To address these challenges, recent work has focused on whether PCCT can overcome these limitations to further improve AC. Hamill et al.~\cite{hamillTechnicalNoteCan2024} investigated whether PCCT can enhance the accuracy of AC in the presence of iodine contrast. They scanned an electron density phantom using both conventional CT and PCCT. The inserts simulated various human tissues, including bone and iodinated blood. In the case of PCCT, VMIs were created at 190 keV. They reported that the LACs derived from conventional CT overestimated the calculated values by up to \SI{18}{\percent} for inserts containing iodine, whereas the corresponding PCCT-based measurements agreed within \SI{3}{\percent}. However, it has also been demonstrated that the use of contrast agents during CT scans can lead to an overestimation of attenuation, resulting in an upward bias in SUV measurements of up to \SI{15}{\percent} \cite{cohadeInitialExperienceOral, dizendorfCauseMagnitudeErrora, mawlawiQuantifyingEffectIV2006}. In general, the use of contrast agents produces artefacts of up to about \SI{20}{\percent} \cite{boellaardStandardsPETImage2009}. Alternatively, VNC and VNI images can be used to virtually remove the iodine contrast from the CT images. However, this requires another transformation scheme, along with a systematic investigation of how it deviates from Carney's method regarding kVp and CT monoenergetic reconstructed energy dependencies. 

In this study, we propose a transformation scheme obtained from a set of measurements across different virtual energies. Using this framework, HUs measured with a PCCT scanner are associated with known \SI{511}{\keV} LACs for a range of reference tissue inserts from the Sun Nuclear electron density phantom \cite{AdvancedElectronDensity,MultiEnergyCTPhantom}, and a bilinear transformation is derived by fitting the reference tissue measurement pairs. We then further evaluated how the resulting conversions, based on VNI and VNC images from contrast-enhanced angiographic CT scans, differ from Carney et al.'s standard method~\cite{carneyMethodTransformingCT2006}, which relies on single polyenergetic CT data. The resulting model is finally compared to Carney's method using data from five patient images. Ultimately, the resulting transformation will be used to develop an AM template derived from spectral CT for our in-house developed BrainPET 7T system, similar to that currently used \cite{rotakopsComparisonTemplateBasedCTBased2015} for a Siemens 3T MR BrainPET insert \cite{caldeiraJulichExperience3T}. 

\section{Materials and Methods}\label{Materials and Methods} 

The Sun Nuclear electron density phantom \cite{AdvancedElectronDensity, MultiEnergyCTPhantom} was scanned using the Siemens Healthineers NAEOTOM Alpha PCCT scanner \cite{siemens_naeotom_alpha}. To generate the AM necessary for PET, a bilinear transformation was derived to map VMI-based HUs to \SI{511}{\keV} LACs. This transformation was subsequently applied to head and neck angiography images from five patient images (three males and two females) to generate their AMs. The study was involved under ethical approval of EK 24-270 (Ethikkommittee der Medizinischen Fakultät, Universität RWTH Aachen). We explored three approaches for obtaining the AMs. In the first approach, VNC and VNI images were combined. The second approach relied only on VNC images, and the third approach used only VNI images. The resulting AMs were compared with the AMs generated using the Carney method, which was applied to native head CT images obtained prior to administration of the iodinated contrast agent~\cite{carneyMethodTransformingCT2006}.

\subsection{Used Phantom}\label{Phantoms}

The Sun Nuclear electron density phantom \cite{AdvancedElectronDensity,MultiEnergyCTPhantom} enables the calibration of tissue-equivalent CT values to electron density. The inserts are manufactured to generate the correct CT image contrasts for human tissue densities, in accordance with ICRU-44 \cite{JournalICRUVolume} and ICRP specifications. The in-plane dimensions of the phantom are \numproduct{40x30} \si{cm^2}, including a removable head section with a \SI{20}{\cm} diameter. This was the only part used in this study, along with the following interchangeable inserts: lung LN-300, lung LN-450, adipose, breast, liver, brain, true water, inner bone, cortical bone, \SI{50}{\percent} \ce{CaCO3}, three blood-mimicking materials with relative electron densities of 1.03, 1.07, and 1.10, and two blood-mimicking materials with iodine content of \SI{2}{\mg/\ml} and \SI{4}{\mg/\ml}. Each insert has a diameter of \SI{28.6}{\mm} and a length of \SI{30}{\cm}. The mass densities and chemical compositions of the inserts are shown in Table~\ref{physical density_Sun Nuclear phantom}, (provided by the manufacturer). 

\subsection{CT Scanner and CT Scan Protocols}\label{Scanners} 

The Siemens Healthineers NAEOTOM Alpha PCCT scanner \cite{siemens_naeotom_alpha} was used to determine the HUs values for each insert. This scanner is a type of spectral CT system \cite{willeminkPhotoncountingCTTechnical2018}, which, among other capabilities, allows for the reconstruction of VMIs. The VMIs are generated by employing the differences in HU values between low- and high-energy datasets to produce images at a specified energy (keV) level. Additionally, PCCT facilitates the reconstruction of VNC and VNI images, both of which are derived from the head and neck angiography images. In VNC images, iodine-associated contrast is fully removed, and calcium-related contrast is partially subtracted, whereas in VNI images, only iodine contrast is fully removed \cite{emrichCoronaryComputedTomography2022a, sodicksonDualEnergyCT2021, VIRARKAR2022293}. 

The Sun Nuclear electron density phantom was scanned using the PCCT scanner with two different protocols. The first protocol was tailored for head and neck angiography, and was performed in spiral mode at tube voltages of \SI{120}{\kV} and \SI{140}{\kV}, with VMIs reconstructed at \SI{64}{\keV}, \SI{67}{\keV}, \SI{70}{\keV}, and \SI{76}{\keV} to obtain the VNI images. In accordance with the manufacturer's recommendations, the VNC images were consistently reconstructed at \SI{70}{\keV}. For the monoenergetic image reconstruction and SPP (Spectral Post-Processing) processing, the Qr40 kernel and Q3 strength level were used at \SI{55}{\keV}.

The second protocol was a native head CT. It involved scanning the phantom in conventional mode (slice-by-slice) at a tube voltage of \SI{120}{\kV}. From this acquisition, VMIs were reconstructed at \SI{67}{\keV}, \SI{70}{\keV}, \SI{80}{\keV}, and \SI{100}{\keV}. Additionally, \SI{3}{\mm} multiplanar reformations (MPR) were generated, and axial images were reconstructed using the Hr36 reconstruction kernel with a Q2 strength level. For the monoenergetic reconstruction and SPP processing, the Qr40 kernel and Q2 strength level were used at \SI{67}{\keV}. No filter was used in either protocol. Beam hardening correction was applied during the measurements, and the reference mAs (milliampere-seconds) (IQ level) was set to 266, leading to an effective current-time product of 291\,\text{eff.\ mAs}.

To evaluate the conversion accuracy in the AM domain, AMs were computed from five patient images from RWTH Aachen University Hospital (three males and two females). All scans were reconstructed with the same protocols and reconstruction parameters as those applied to the phantom scans, and all patient data were fully anonymized. For our study, we used spectral data sets corresponding to head and neck angiography at \SI{120}{\kV} (for three patient images) and at \SI{140}{\kV} (for two patient images), with previous administration of an iodinated contrast agent to compute VNC and VNI images. In addition, contrast agent-free (native) head CT images acquired at \SI{120}{\kV} for the same patients and reconstructed at \SI{70}{\keV} were used to compare the VNC and VNI-based AMs with those obtained using Carney's method.

\subsection{Transformation scheme from VNC/VNI image Hounsfield Units to attenuation correction factors in PET}\label{Transformation Scheme}

\begin{figure}[ht!] 
       \centering
       \begin{subfigure}{0.45\textwidth}
           \includegraphics[width=\textwidth]{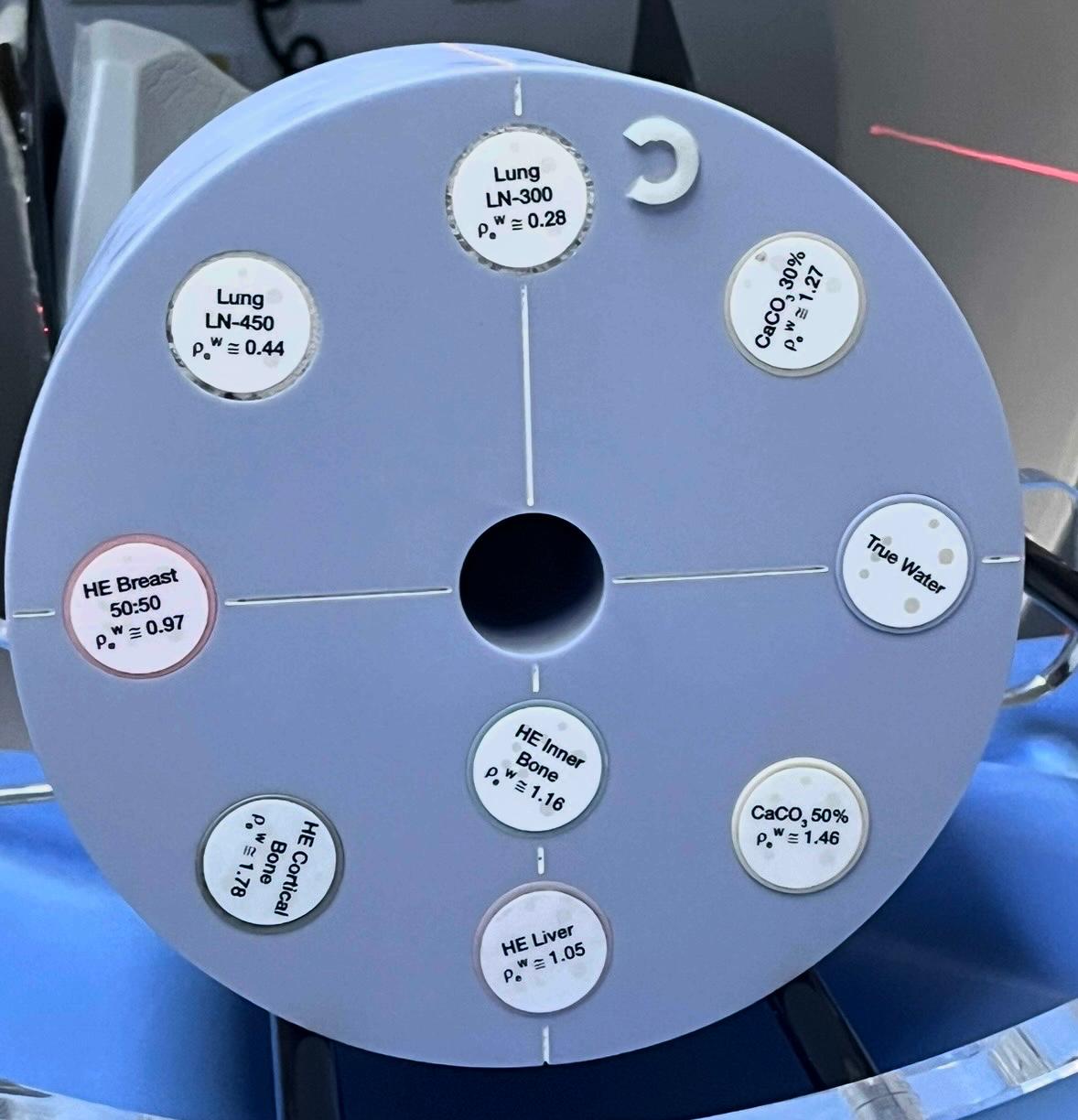} 
           \caption{Set 1}
           \label{fig:set_1}
       \end{subfigure}
       \hfill
       \begin{subfigure}{0.45\textwidth}
           \includegraphics[width=\textwidth]{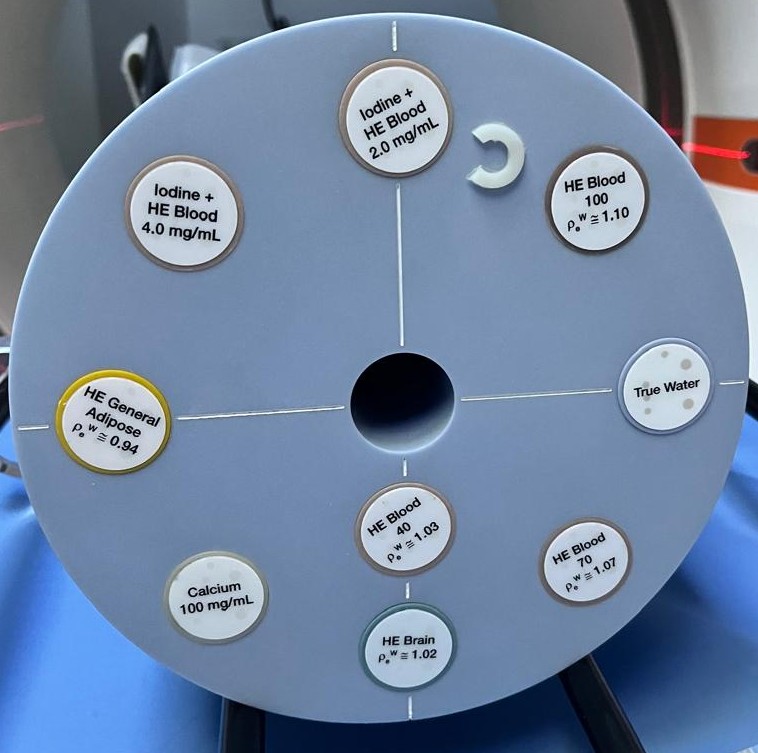} 
           \caption{Set 2}
           \label{fig:set_2}
       \end{subfigure}
       \caption{Insert layouts for Set 1 and Set 2 for the Sun Nuclear electron density phantom. For Set 1, the inserts—arranged clockwise starting from the top—are: lung LN-300, \SI{30}{\percent} \ce{CaCO3}, true water, \SI{50}{\percent} \ce{CaCO3}, liver, cortical bone, breast, lung LN-450, and inner bone. For Set 2, the inserts—also arranged clockwise from the top—are: blood + iodine \SI{2}{\mg/\ml}, HE blood 100, true water, HE blood 70, brain, calcium \SI{100}{\mg/\ml}, adipose, blood + iodine \SI{4}{\mg/\ml}, and HE blood 40. A glued flat washer was used to facilitate the identification of the inserts in Amide. \it Not all of the inserts shown were used in this study, i.e., the inserts 30\% \ce{CaCO3}, true water, and calcium \SI{100}{\mg/\ml} were only included in the phantom during CT scanning to avoid potential image artefacts caused by large air cavities. The inserts included in this study are listed in Table~\ref{physical density_Sun Nuclear phantom}.}
       \label{Sun Nuclear electron density phantom}
\end{figure}

To measure the HUs, the phantom was scanned using the PCCT, as described in~\ref{Scanners}. As the removable head section of the phantom could only accommodate ten inserts, but a total of fifteen were utilized in this study, two scans were performed, using separate sets (Set 1 and Set 2). The high-density tissue replacement rod inserts (e.g., bone-equivalent materials) were uniformly distributed throughout the phantom to minimize potential reconstruction artefacts. The remaining cavities in the phantom were filled with other inserts that were not analysed in this study. The configurations of Sets 1 and 2 are illustrated in Figure~\ref{Sun Nuclear electron density phantom}.

\begin{figure}[ht!]
    \centering
    \includegraphics[width=0.5\textwidth]{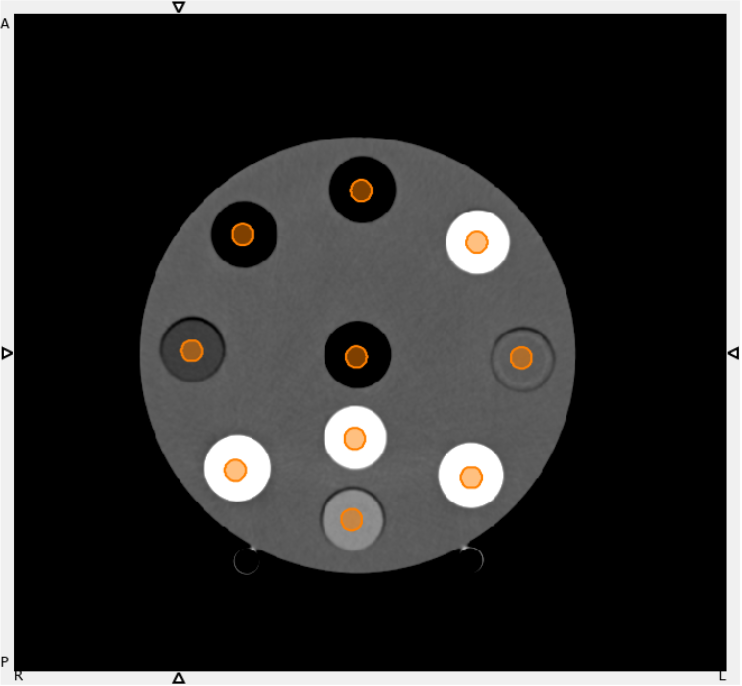}
    \caption{Cylindrical ROIs with a diameter of \SI{10}{\mm} and a length of \SI{10}{\mm}, positioned centrally within each insert in the PCCT image}
    \label{ROI_PCCT}
\end{figure}

Amide (version 1.0.6) was used to extract HUs from the phantom CT images \cite{AMIDESoftware2003}. Cylindrical regions of interest (ROIs) were defined with a diameter of \SI{10}{\mm} and a length of \SI{10}{\mm} and positioned centrally within each insert. The HUs were then extracted from these ROIs for each insert using the ROI statistics tools in Amide. The ROIs with the PCCT images are shown in Figure~\ref{ROI_PCCT} for one scan. The standard error (SE) of the measured insert HUs was calculated using $\mathrm{SE}=\sigma/\sqrt{N}$, where $\sigma$ is the standard deviation, and $N$ is the number of the voxels in the ROIs.

The XCOM database provided by the National Institute of Standards and Technology (NIST) was used to determine the theoretical LACs at \SI{511}{\keV} for the phantom inserts \cite{NISTdatabase}. The calculations were performed using the chemical compositions provided in the Sun Nuclear electron density phantom data sheets. The total mass attenuation coefficients with coherent scattering were used at 511 keV and were subsequently multiplied by the corresponding mass densities to obtain the LACs. The values are shown in Table~\ref{physical density_Sun Nuclear phantom}. According to Report 44 of the International Commission on Radiation Units and Measurements \cite{JournalICRUVolume}, tissue-equivalent materials should have LACs within $\pm$\SI{5}{\percent} of those of human tissue over the relevant photon energy range. The relative differences between LACs derived from the insert compositions and measured inserts are within $\pm$\SI{3}{\percent} for all relevant head tissue types, as reported in our previous study \cite{OurPreviousStudy}.

Bilinear transformation formulas were obtained by regression with two linear models, which are joined at a defined break point (BP) corresponding to the HU value at which the two lines intersect, estimated using R (version 4.3.3). For each insert, the HU values obtained from the PCCT reconstructions were paired with the reference LACs derived from the NIST XCOM database, which are listed in Table~\ref{physical density_Sun Nuclear phantom}. To assess the uncertainty of the derived parameters, a linear regression was performed on the inverted dependency, i.e.~taking the LAC values as the $x$ values and the HU values and their uncertainties as the $y$ values, along with their corresponding errors. After the regressions, the best-fit linear functions obtained were inverted back, and Gaussian error propagation was used to compute the slope and intercept uncertainties for the inverted linear function from the fit parameter uncertainties \cite{press1992numerical}. Using this approach, the slope and intercept of the inverted back functions are $1/b$ and $-a/b$, respectively. Additionally, the errors in the slope and intercept of the linear relationships with uncertainties in the x-coordinates are given by equation~\ref{error propagation equation}, where $a$ is the intercept of the inverted functions and $b$ is the slope of the inverted functions, their uncertainties are represented by $\sigma_{a}$ and $\sigma_{b}$, respectively.

\begin{equation}
\begin{aligned}
\sigma_{-a/b}^{2}
&=
\left( \frac{-1}{b} \right)^{2} \sigma_{a}^{2}
+
\biggl( \frac{a}{b^{2}} \biggr)^{2} \sigma_{b}^{2}
+
2 \left( \frac{-1}{b} \right)
  \biggl( \frac{a}{b^{2}} \biggr)
  \operatorname{Cov}(a,b)
\\
\sigma_{1/b} 
&= 
\frac{\sigma_{b}}{b^{2}}
\end{aligned}
\label{error propagation equation}
\end{equation}

\begin{table}[ht!]
\caption{Mass density $\rho$ (\si[per-mode=single-symbol]{\g\per\cm\cubed}), chemical composition, and insert composition LACs from XCOM at \SI{511}{\keV} of the used inserts of the Sun Nuclear electron density phantom}\label{physical density_Sun Nuclear phantom}
\begin{tabular}{ p{11mm} p{4mm} p{4mm} p{4mm} p{4mm} p{4mm} p{22mm} p{13mm} p{18mm}}
\toprule
Inserts                            &   H  &   B  &    C  &   N  &   O   & Others                                                  & $\rho$ (\si[per-mode=single-symbol]{\g\per\cm\cubed}) & LACs (\si[per-mode=symbol]{\per\cm}) \\
\midrule
Liver                              & 8.25 & 0.03 & 66.87 & 2.25 & 19.02 & 0.10 Na, 0.75 Mg, 0.65 Si, 0.14 Cl, 1.94 Ca             & 1.08 &  0.10087\\
Brain                              & 8.23 & 0.04 & 65.75 & 2.05 & 19.70 & 0.15 Na, 1.23 Mg, 0.94 Si, 0.13 Cl, 1.79 Ca             & 1.05 &  0.09805\\
Breast                             & 9.48 & 0.04 & 70.23 & 2.47 & 15.13 & 0.16 Na, 0.61 Mg, 1.01 Si, 0.12 Cl, 0.74 Ca             & 0.98 &  0.09254\\
Adipose                            & 9.73 & 0.05 & 71.41 & 2.71 & 14.35 & 0.18 Na, 1.11 Si, 0.12 Cl, 0.34 Ca                      & 0.96  &  0.09085\\
Lung LN-300                        & 7.43 &      & 57.86 & 1.96 & 20.71 & 11.19 Mg, 0.77 Si, 0.08 Cl                              & 0.29   &   0.02686\\
Lung LN-450                        & 7.44 &      & 58.03 & 1.97 & 20.69 & 11.22 Mg, 0.57 Si, 0.08 Cl                              & 0.45   &   0.04168\\
Inner bone                         & 6.38 & 0.03 & 53.79 & 1.73 & 25.64 & 0.11 Na, 1.68 Mg, 0.72 Si, 0.10 Cl, 9.82 Ca             & 1.21  &  0.11122\\
Cortical bone                      & 2.31 &      & 27.41 & 0.85 & 39.95 & 2.82 Mg, 0.04 Cl, 26.62 Ca                              & 1.93   &  0.17119 \\
50\% \ce{CaCO3}                    & 4.03 &      & 40.34 & 1.52 & 34.13 & 0.30 Mg, 0.07 Cl, 19.62 Ca                              & 1.56 &  0.14052 \\
HE Blood 40                        & 8.37 & 0.03 & 66.85 & 2.16 & 18.59 & 0.13 Na, 1.11 Mg, 0.79 Si, 0.13 Cl, 1.74 Ca, 0.10 Fe    & 1.06 &  0.09911\\
HE Blood 70                        & 8.44 & 0.02 & 67.42 & 2.18 & 18.35 & 0.08 Na, 1.11 Mg, 0.49 Si, 0.14 Cl, 1.67 Ca, 0.10 Fe    & 1.095  &  0.10244\\
HE Blood 100                       & 8.50 & 0.01 & 67.96 & 2.20 & 18.12 & 0.03 Na, 1.11 Mg, 0.22 Si, 0.14 Cl, 1.61 Ca, 0.10 Fe    & 1.13  &  0.10578\\
Blood 40 + \SI{2}{\mg/\ml} Iod     & 8.36 & 0.03 & 66.75 & 2.15 & 18.53 & 0.12 Na, 1.11 Mg, 0.78 Si, 0.13 Cl, 1.73 Ca, 0.10 Fe, 0.19 I     & 1.061 &  0.09921  \\
Blood 40 + \SI{4}{\mg/\ml} Iod     & 8.34 & 0.03 & 66.65 & 2.15 & 18.47 & 0.12 Na, 1.10 Mg, 0.78 Si, 0.13 Cl, 1.73 Ca, 0.10 Fe, 0.39 I     & 1.063 &  0.09939  \\
\botrule
\end{tabular}
\end{table}

\subsection{Obtaining the attenuation map from virtual monoenergetic images}\label{Obtaining the Attenuation Map from Virtual Monoenergetic Images}

To obtain the AMs of the five patient images using the derived bilinear transformation described above, three approaches were evaluated. In the first approach, both the VNC and VNI images were used in combination. The second approach relied solely on VNC images, while the third approach used only VNI images. For this study, VMI images at \SI{70}{\keV} were chosen, since the VNC images are consistently generated at this energy. This consistency enabled reliable combination with VNI images reconstructed at the same energy and facilitated direct comparisons. Both the VNC and VNI images were registered to the native head CT scans acquired without iodine contrast agent administration, allowing comparison of the AMs derived from CT images with iodinated blood to the standard method from Carney et al.~\cite{carneyMethodTransformingCT2006}. The registration to the native head CT was performed using rigid registration with the ANTs Python package, ANTsPy (version 0.6.1) \cite{avantsReproducibleEvaluationANTs2011}.

\begin{figure}[ht!]
    \centering
    \includegraphics[width=1\textwidth]{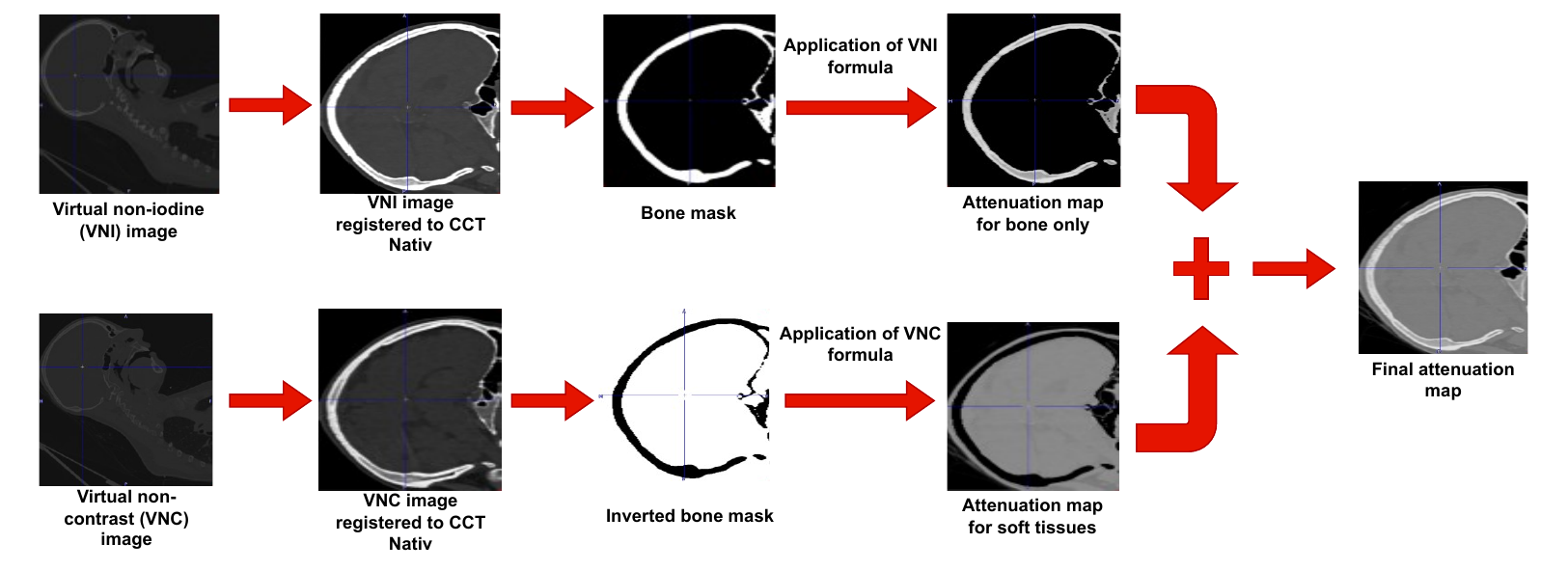}
    \caption{Workflow for generating the combined attenuation map from VNC and VNI images}
    \label{our-approach-steps}
\end{figure}

In the first approach, a mask of the VNI patient brain image was generated using the BP as a threshold value of \SI{118}{HU} for patient images scanned at \SI{120}{\kV} and \SI{231}{HU} for images scanned at \SI{140}{\kV}. These BP values were derived from the VNC transformation scheme by computing the intersection point. Subsequently, HU values were converted to the LACs at \SI{511}{\keV} using the VNI-based bilinear equation for the segmented bone voxels and using the VNC-based bilinear equation for all other tissue voxels. Finally, the LAC values for the bone voxels derived from the VNI image were combined with the LAC values for all other voxels derived from the VNC image to generate a complete AM for each patient. The procedure is illustrated in Figure~\ref{our-approach-steps}. 

In the second approach, we applied the corresponding VNC-based bilinear equation for all values above and below the BP. In the third approach, the corresponding VNI-based bilinear equation was applied for all values above and below the BP. For all three approaches, the corresponding equations are given in detail in section~\ref{Results}. All processing steps were implemented in Python (version 3.9.13). NumPy (version 2.2.6) and NiBabel (version 5.3.2) packages were used. 

\subsection{Comparison to Carney's transformation}\label{Validation in patients in data analysis}

For comparison to Carney's method \cite{carneyMethodTransformingCT2006}, native head CT images acquired at \SI{120}{\kV} and \SI{70}{\keV} before the administration of the iodinated contrast agent were used. As the Carney method provides kVp-dependent parameters (\SI{80}{\kV}, \SI{100}{\kV}, \SI{110}{\kV}, \SI{120}{\kV}, \SI{130}{\kV}, and \SI{140}{\kV}) but not virtual reconstruction energy, we estimated the corresponding mean spectrum energies using SpekCalc \cite{poludniowskiSpekCalcProgramCalculate2009}, yielding \SI{46.3}{\keV}, \SI{52.5}{\keV}, \SI{55.2}{\keV}, \SI{57.7}{\keV}, \SI{60}{\keV}, and \SI{62.1}{\keV}, respectively. The closest match to the native head CT images at \SI{70}{\keV} is the mean spectrum energy for \SI{140}{\kV}. The parameters for the \SI{140}{\kV} were selected according to the Carney equation \cite{carneyMethodTransformingCT2006}. The equations used are given in Equation~\ref{carney_equ}. The BP for the bilinear transformation was \SI{30}{HU}.

\begin{align}
\mathrm{Below\text{ }Break\text{ }Point}{:}\text{ }\mu &= 9.6\times10^{-5}(\si{HU} + 1000) \nonumber \\
\mathrm{Above\text{ }Break\text{ }Point}{:}\text{ }\mu &= 5.64\times10^{-5}(\si{HU} + 1000) + 4.08\times10^{-2} 
\label{carney_equ}
\end{align}

A comparative analysis was performed between the two AMs using all five patient images and the following tissue classes: soft tissue, bone, mastoid cells, and blood vessels. For the areas corresponding to soft tissue, bone, and mastoid cell regions, the non-head background, i.e.~imaged areas that are not part of the human head, was removed using PMOD (PMOD Technologies LLC, Zürich, Switzerland, version 4.303). Segmentation was performed with the Otsu thresholding tool using 20 histogram bins, a threshold of 2, and 1 cluster \cite{Otsu}. These parameters were selected to provide robust brain masking while minimizing the inclusion of non-brain structures and noise. The resulting mask was subsequently applied to both AMs for each patient image using Python. Voxel classification thresholds were as follows: $ \mu <  \SI[per-mode=symbol]{0.085}{\per\cm} $ for mastoid cells or mixed tissue, $ \SI[per-mode=symbol]{0.085}{\per\cm} \leq \mu \leq \SI[per-mode=symbol]{0.105}{\per\cm} $ for soft tissue, and $ \mu > \SI[per-mode=symbol]{0.105}{\per\cm} $ for bone \cite{kopsAttenuationCorrectionCerebellum2024}.

For the comparison focusing solely on the effects within iodinated blood, head and neck angiography images, i.e.~the images acquired after administration of the iodine contrast agent, were registered to the native CT scans for each patient image. Tubular structures corresponding to the blood vessels were selected by applying a multiscale Frangi vesselness filter to the co-registered angiographic images \cite{frangiMultiscaleVesselEnhancement1998}, as implemented in the Python package scikit-image (version 0.25.2). Small isolated components were removed, and morphological operations—including erosion, dilation, hole filling, and closing—were applied to smooth the vessel boundaries and eliminate thin inter-connections. The resulting vessel ROI masks were saved for subsequent quantitative comparison.

\subsection{Data Analysis}\label{performance precision in data analysis}

The precision between both methods with respect to the generated AMs was assessed by computing the Dice coefficients (DC), sensitivity, specificity, and accuracy for soft tissue, bone, and mastoid cell regions. For each subject and AM, the DC values between the proposed AM and the reference Carney AM were calculated as defined in Equation~\ref{Dice coefficients equation}.  

\begin{equation}
Dice = 2 \times \frac{\sum (\text{VOX}_{\mathrm{Carney AM}} \cap \text{VOX}_{\mathrm{Proposed AM}})}
{\sum (\text{VOX}_{\mathrm{Carney AM}}) + \sum (\text{VOX}_{\mathrm{Proposed AM}})}
\label{Dice coefficients equation}
\end{equation}

In addition, the relative differences (RD) between the two methods were calculated as described in Equation~\ref{relative differences equation} and the Root Mean Squared Deviation (RMSD), given in Equation~\ref{RMSD equation}, was computed for the soft tissue, bone, mastoid cell, and blood vessel regions. The minimum and maximum values of the difference images, representing the differences between the AMs obtained using the Carney method and the proposed transformation scheme, were also identified and reported. All metrics were computed using Python, and the backgrounds (non-head structures) were removed as explained above. In all equations, \text{VOX} represents the voxel values.

\begin{equation}
\mathrm{RD} = \frac{1}{N} \sum_{x=1}^{N}  \frac{\text{VOX}_{\mathrm{Proposed AM}} - \text{VOX}_{\mathrm{Carney AM}}} {\text{VOX}_{\mathrm{Carney AM}}} \times \SI{100}{\percent}
\label{relative differences equation}
\end{equation}

\begin{equation}
\mathrm{RMSD} = \sqrt{ \frac{1}{N} \sum_{x=1}^{N} \left( \text{VOX}_{\mathrm{Proposed AM}} - \text{VOX}_{\mathrm{Carney AM}} \right)^2 }
\label{RMSD equation}
\end{equation}

\section{Results}\label{Results}

\begin{figure}[ht!] 
       \centering
       \begin{subfigure}{0.45\textwidth}
           \includegraphics[width=\textwidth]{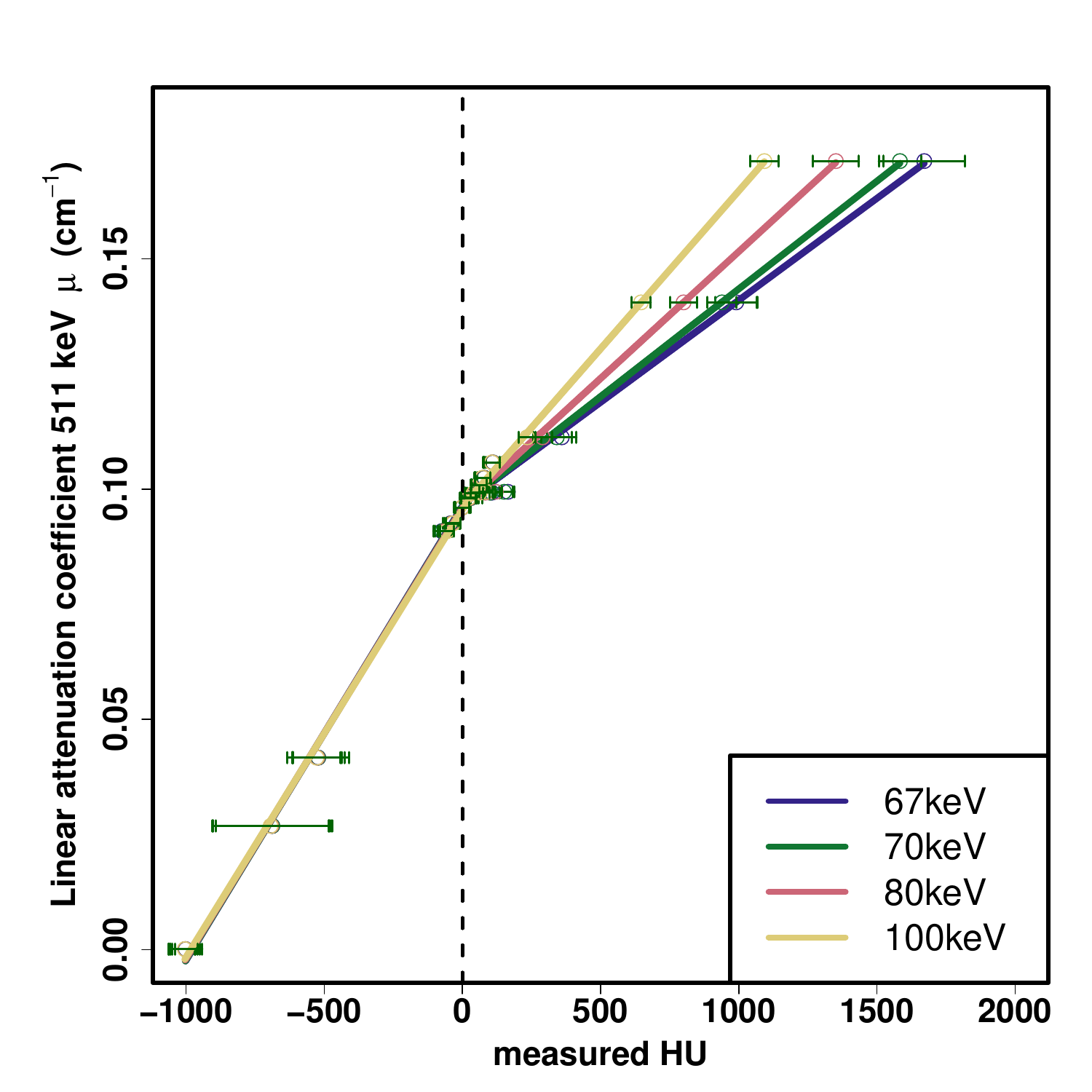} 
           \caption{}
       \end{subfigure}
       \hfill
       \begin{subfigure}{0.45\textwidth}
           \includegraphics[width=\textwidth]{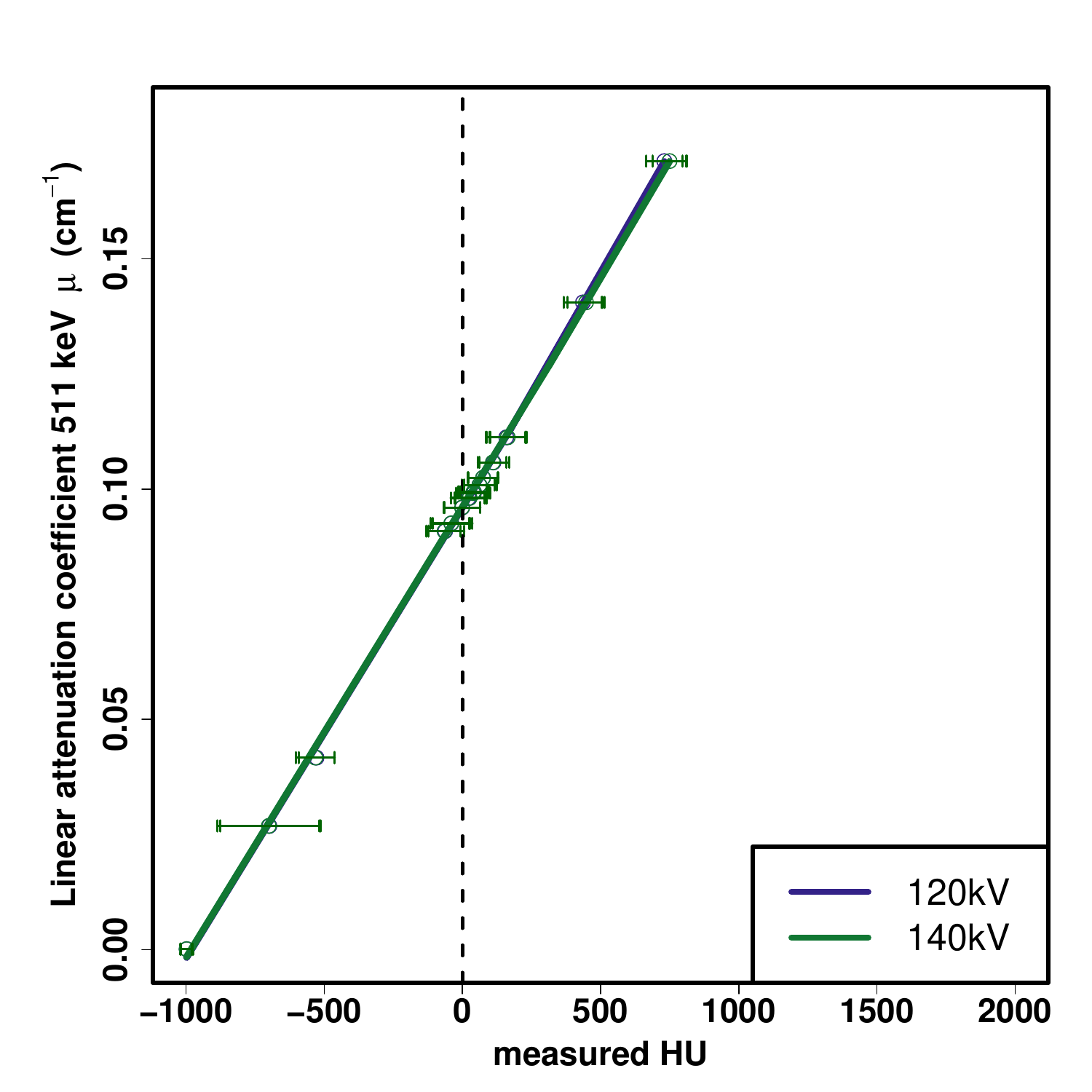} 
           \caption{}
       \end{subfigure}
       
       \begin{subfigure}{0.45\textwidth}
           \includegraphics[width=\textwidth]{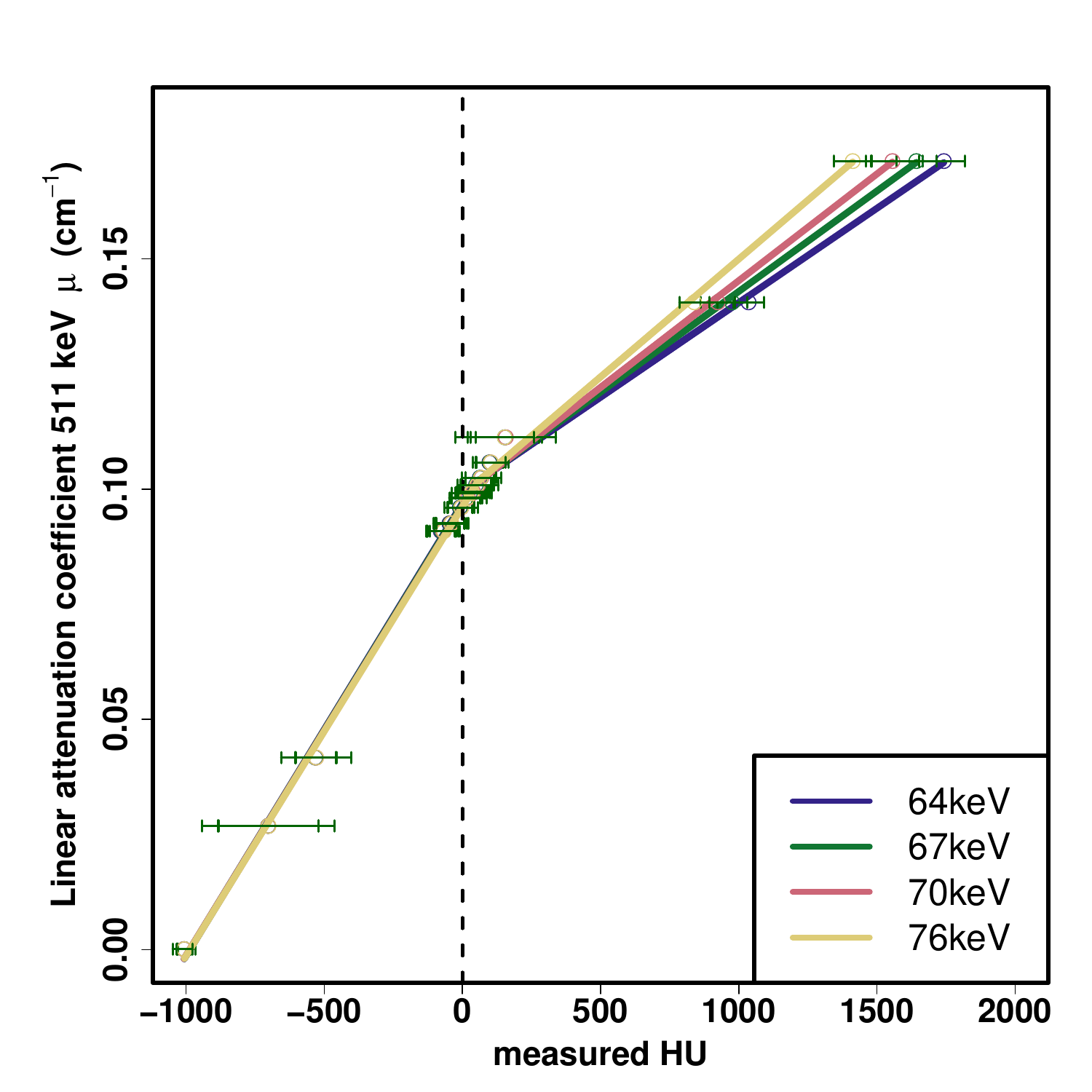} 
           \caption{}
       \end{subfigure} 
       \hfill
       \begin{subfigure}{0.45\textwidth}
           \includegraphics[width=\textwidth]{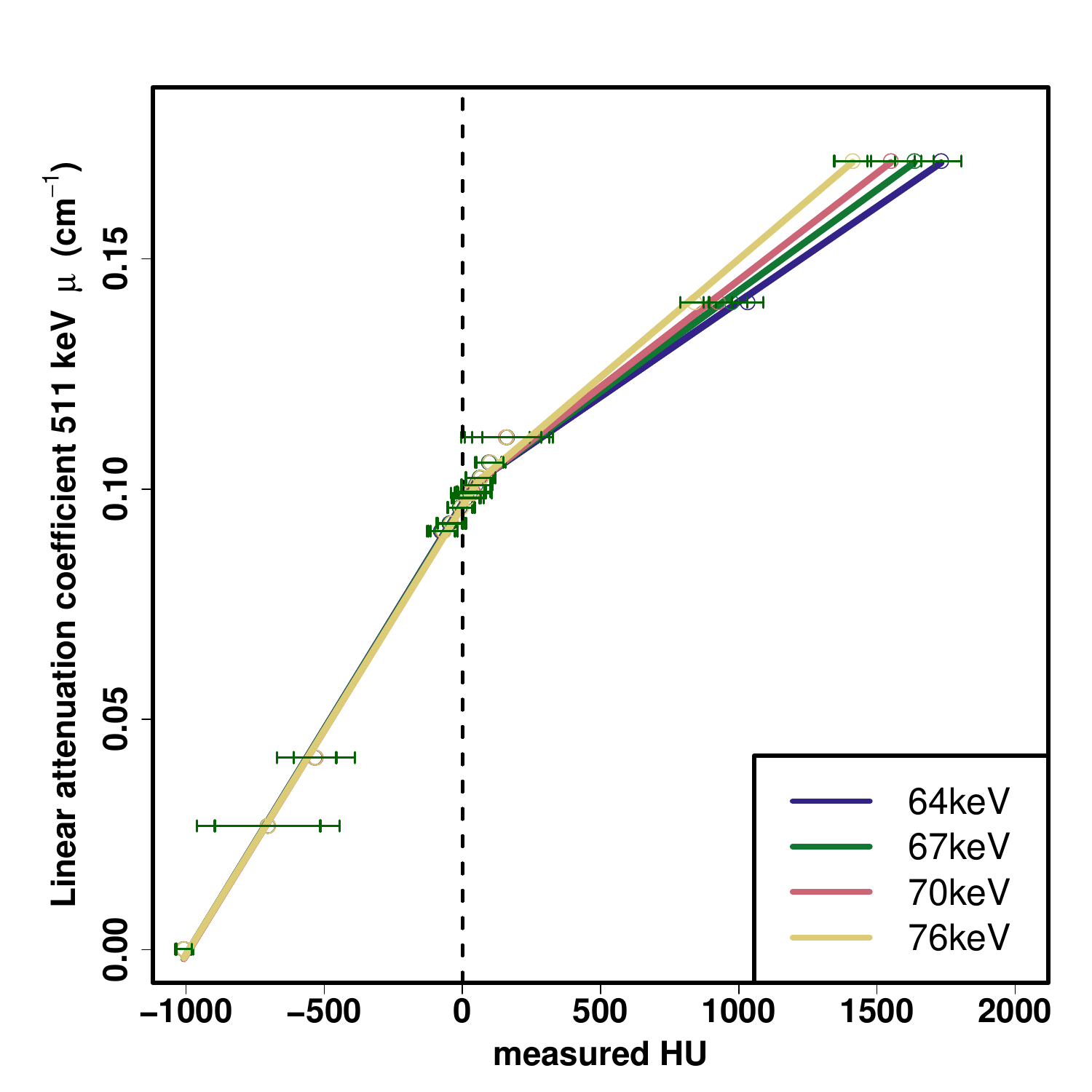} 
           \caption{}
       \end{subfigure} 
       
       \caption{Monoenergetic linear attenuation values for the inserts in the Sun Nuclear electron density phantom vs.~HU a) native head CT at \SI{120}{\kV} tube voltage for virtual monoenergetic CT images at energies of \SI{67}{\keV}, \SI{70}{\keV}, \SI{80}{\keV}, and \SI{100}{\keV} b) VNC at \SI{120}{\kV} and \SI{140}{\kV} tube voltage c) VNI at \SI{120}{\kV} tube voltage for VMIs at \SI{64}{\keV}, \SI{67}{\keV}, \SI{70}{\keV}, and \SI{76}{\keV} d) VNI at \SI{140}{\kV} tube voltage for VMIs at \SI{64}{\keV}, \SI{67}{\keV}, \SI{70}{\keV}, and \SI{76}{\keV}. Note: the error bars representing the standard error of the mean for the HU values were multiplied by 600 to aid visibility}
       \label{Extrapolation schemes}
 \end{figure}

The conversion schemes were computed for tube voltages of \SI{120}{\kV} and \SI{140}{\kV}, and energies of \SI{64}{\keV}, \SI{67}{\keV}, \SI{70}{\keV}, \SI{76}{\keV}, \SI{80}{\keV}, and \SI{100}{\keV}. Figure~\ref{Extrapolation schemes} shows the dependencies of the LACs on the HUs for different CT contrasts and at different VMI energies. 

Equation~\ref{VNI_equ} and Table~\ref{VNI_table} present the obtained transformation for tissues based on the VNI images. Equation~\ref{VNC_equ} corresponds to the VNC images which is, as expected, almost identical for \SI{120}{\kV} and \SI{140}{\kV}, as the VNC images are consistently reconstructed at \SI{70}{\keV}. For VNC equations, the slope and intercept were calculated by combining the \SI{120}{\kV} and \SI{140}{\kV} estimates with the weighted average of two independent fits. For each parameter \(p\) with uncertainties (standard error, SE) \(\sigma_p\), the weight was defined as \(w = 1/\sigma_p^2\), and the combined value was computed as \(p_{\mathrm{mid}} = (p_1 w_1 + p_2 w_2)/(w_1 + w_2)\). The corresponding uncertainty was then taken as \(\sigma_{\mathrm{mid}} = 1/\sqrt{w_1 + w_2}\) \cite{hartung2008statistical}. The BPs for equation~\ref{VNC_equ} are 118.14 HU and 231.34 HU for \SI{120}{\kV} and \SI{140}{\kV}, respectively. All values reported in the equations and the table are given with their SE.

\begin{align}
\mu &= a \times \si{HU} + b 
\label{VNI_equ}
\end{align}

\begin{align}
\mathrm{Below\text{ }Break\text{ }Point}{:}\text{ } \mu &= ( 9.63 \pm 0.08)\times10^{-5}\times \si{HU} + (9.60 \pm 0.07)\times10^{-2} \nonumber \\
\mathrm{Above\text{ }Break\text{ }Point}{:}\text{ } \mu &= (10.25 \pm 0.04)\times10^{-5}\times \si{HU} + (9.49 \pm 0.01)\times10^{-2}
\label{VNC_equ}
\end{align}

\begin{table}[ht!]
\caption{keV and VMI energy-dependent best-fit parameters $a$ and $b$ together with the corresponding break points (BP) defining the bilinear transformations for Equation~\ref{VNI_equ}, $a$ and $b$ are in units of (\si[per-mode=symbol]{\per\cm})}
\label{VNI_table}
\centering
\begin{tabular*}{\textwidth}{p{1.0cm}p{1.75cm}p{1.6cm}cccccc}
\toprule
\multirow{2}{1cm}{Energy (\si{\keV})} & \multirow{2}{1.75cm}{Tube~Voltage (\si{\kV})} & 
\multirow{2}{1.6cm}{Break Point (BP) (\si{HU})} & \multicolumn{2}{c}{Above break point} & \multicolumn{2}{c}{Below break point} \\
\cmidrule(lr{.05cm}){4-5}\cmidrule(lr{.05cm}){6-7}
& & & $a \times 10^{-5}$ & $b \times 10^{-2}$ & $a \times 10^{-5}$ & $b \times 10^{-2}$ \\
\midrule
\text{64} 
& 120 & 31.78 & $4.14 \pm 0.12$ & $9.89 \pm 0.06$ & $9.66 \pm 0.12$ & $9.71 \pm 0.09$ \\
& 140 & 30.93 & $4.17 \pm 0.12$ & $9.89 \pm 0.06$ & $9.65 \pm 0.11$ & $9.72 \pm 0.08$ \\
\midrule
\text{67} 
& 120 & 35.07 & $4.40 \pm 0.12$ & $9.87 \pm 0.06$ & $9.64 \pm 0.12$ & $9.69 \pm 0.09$ \\
& 140 & 33.50 & $4.42 \pm 0.12$ & $9.87 \pm 0.06$ & $9.63 \pm 0.11$ & $9.69 \pm 0.08$ \\
\midrule
\text{70} 
& 120 & 31.74 & $4.67 \pm 0.11$ & $9.82 \pm 0.06$ & $9.62 \pm 0.11$ & $9.67 \pm 0.07$ \\
& 140 & 39.69 & $4.65 \pm 0.14$ & $9.87 \pm 0.07$ & $9.61 \pm 0.08$ & $9.67 \pm 0.06$ \\
\midrule
\text{76} 
& 120 & 43.71 & $5.15 \pm 0.12$ & $9.83 \pm 0.06$ & $9.60 \pm 0.12$ & $9.63 \pm 0.09$ \\
& 140 & 40.89 & $5.15 \pm 0.11$ & $9.82 \pm 0.05$ & $9.59 \pm 0.11$ & $9.64 \pm 0.08$ \\
\bottomrule
\end{tabular*}
\end{table}

In Figure~\ref{Extrapolation schemes} (b), it can be observed that the VNC-only transformation scheme apparently converges for all cases to a single linear function (although it is reported here as two linear functions with a break point), leading  to a mapping between HU values and LACs for soft tissues which is significantly less ambiguous than the transformation scheme from Carney. Therefore, for this scheme, the absolute and relative differences between the LACs of the insert compositions and the predictions made by the transformation scheme at \SI{120}{\kV} were computed and organized for the different inserts in Table~\ref{relative_residuals_VNC_120kV}. The differences for the VNC transformation scheme at \SI{140}{\kV} are not shown, as they are very similar to those observed at \SI{120}{\kV}. 

For HU values close to and below the value of water (around \SI{0}{HU}), a nearly identical behavior is observed in all cases (see Figure~\ref{Extrapolation schemes}). At higher values, corresponding to denser bone tissue, increasing discrepancies are observed between the measured HU at different \si{\keV} settings. 

\begin{table}[ht!]
\caption{Absolute and relative differences in LACs between computation (theoretical) from XCOM and inserts' composition and prediction by the VNC transformation scheme at \SI{120}{\kV}}\label{relative_residuals_VNC_120kV}
\begin{tabular}{ p{33mm} p{17mm} p{17mm}  p{14mm} p{14mm} }
\toprule
Inserts          &  Theoretical LACs (\si[per-mode=symbol]{\per\cm}) &   Predicted LACs (\si[per-mode=symbol]{\per\cm})   &   Absolute differences ($\times 10^{-4}$) (\si[per-mode=symbol]{\per\cm})  &   Relative differences (\%)   \\
\midrule
Air                                & 0.00010 & -0.00016  & $2.6$  & -165.8 \\
Liver                              & 0.10087 & 0.10121  & $-3.4$   & -0.3 \\ 
Brain                              & 0.09805 & 0.09712  & $9.3$  & 1.0 \\
Breast                             & 0.09254 & 0.09198   & $5.6$  & 0.6 \\
Adipose                            & 0.09085  & 0.08978  & $10.7$  & 1.2 \\
Lung LN-300                        & 0.02686 & 0.02863  & $-17.7$ & -6.2 \\
Lung LN-450                        & 0.04168 &  0.04501     & $-33.3$  & -7.4 \\
Inner bone                         & 0.11122 & 0.11139  & $-1.7$   & -0.2 \\
Cortical bone                      & 0.17119  &  0.17111  & $0.8$  & 0.0  \\
50\% \ce{CaCO3}                    & 0.14052 &  0.14037   & $1.5$  & 0.1 \\
HE Blood 40                        & 0.09911 & 0.09892 & $1.9$  & 0.2 \\
HE Blood 70                        & 0.10244 & 0.10263  &  $-1.9$  & -0.2\\
HE Blood 100                       & 0.10578 & 0.10655  & $-7.7$  &  -0.7 \\
Blood 40 + \SI{2}{\mg/\ml} Iod     & 0.09921 & 0.09920  & $0.1$  & 0.0  \\
Blood 40 + \SI{4}{\mg/\ml} Iod     & 0.09939 & 0.09911  & $2.8$  &  0.3  \\
\botrule
\end{tabular}
\end{table}
  
\subsection{Comparison to Carney's method}\label{Validation in patients in results}

\begin{figure}[ht!]
    \centering
    \includegraphics[width=1\textwidth]{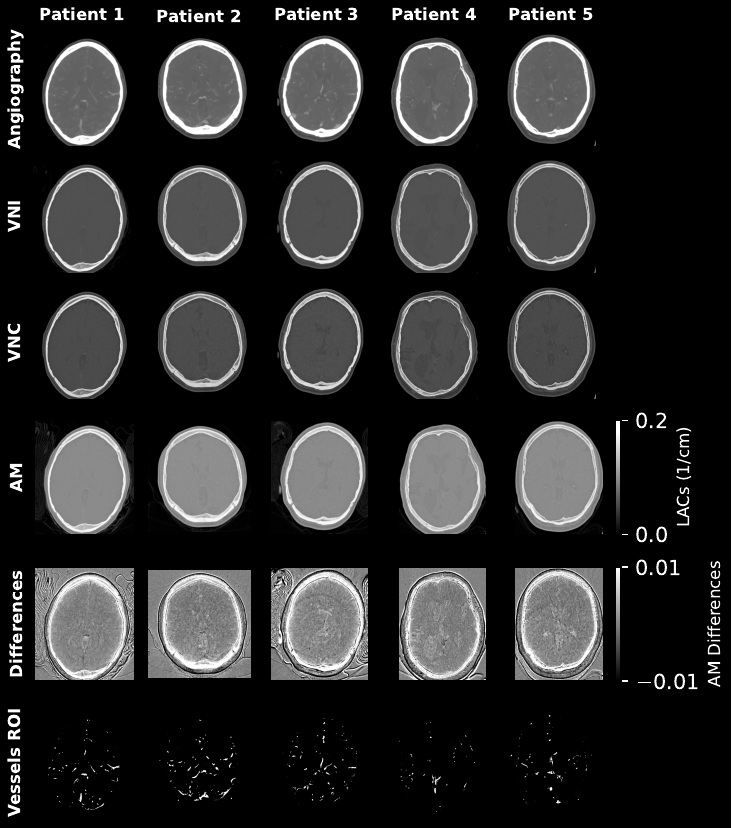}
    \caption{From upper row to lower row: angiographic images, virtual non-iodine images, virtual non-contrast images, attenuation maps generated by the first proposed approach (using both VNC and VNI images), difference images between AMs obtained with the first approach and Carney's method, and the segmented blood vessel ROI}
    \label{Scans_Patients}
\end{figure}

The difference images between AMs from Carney's method and the AMs generated with the first proposed approach are shown in Figure~\ref{Scans_Patients} for the five patients along with the AMs from the first approach, the angiographic images, VNC, VNI, and the segmented blood vessel ROI. 

\subsubsection{Evaluation of overlap}\label{performance precision in results}

\begin{figure}[ht!] 
       \centering
       \begin{subfigure}{1.0\textwidth}
           \includegraphics[width=\textwidth]{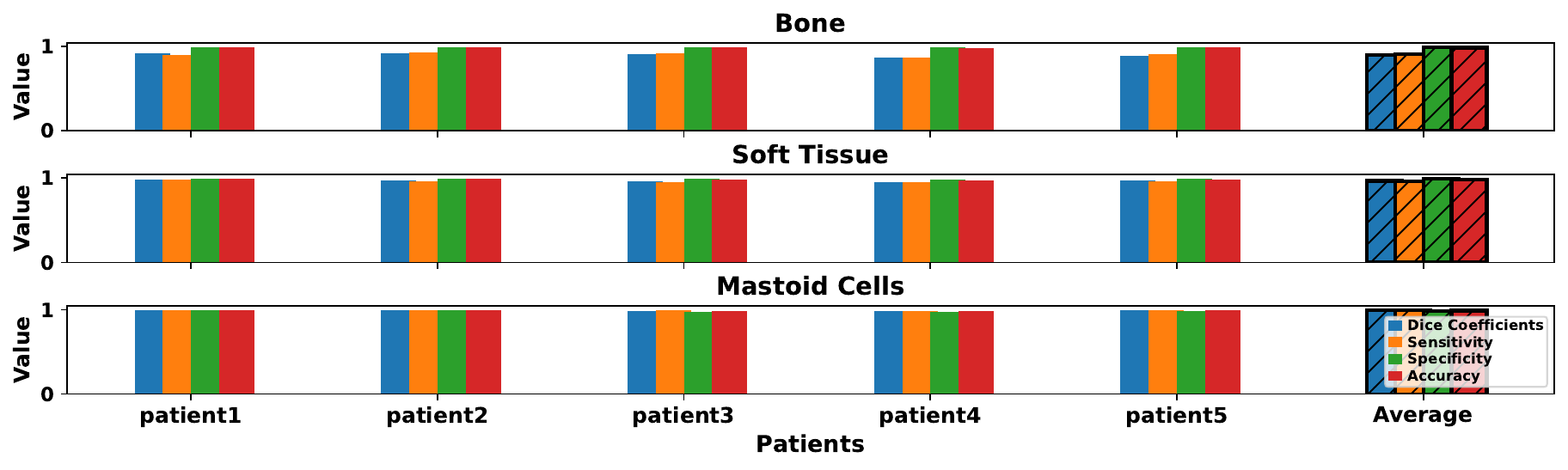} 
           \caption{First approach}
           \label{fig:a}
       \end{subfigure}
       \hfill
       \begin{subfigure}{1.0\textwidth}
           \includegraphics[width=\textwidth]{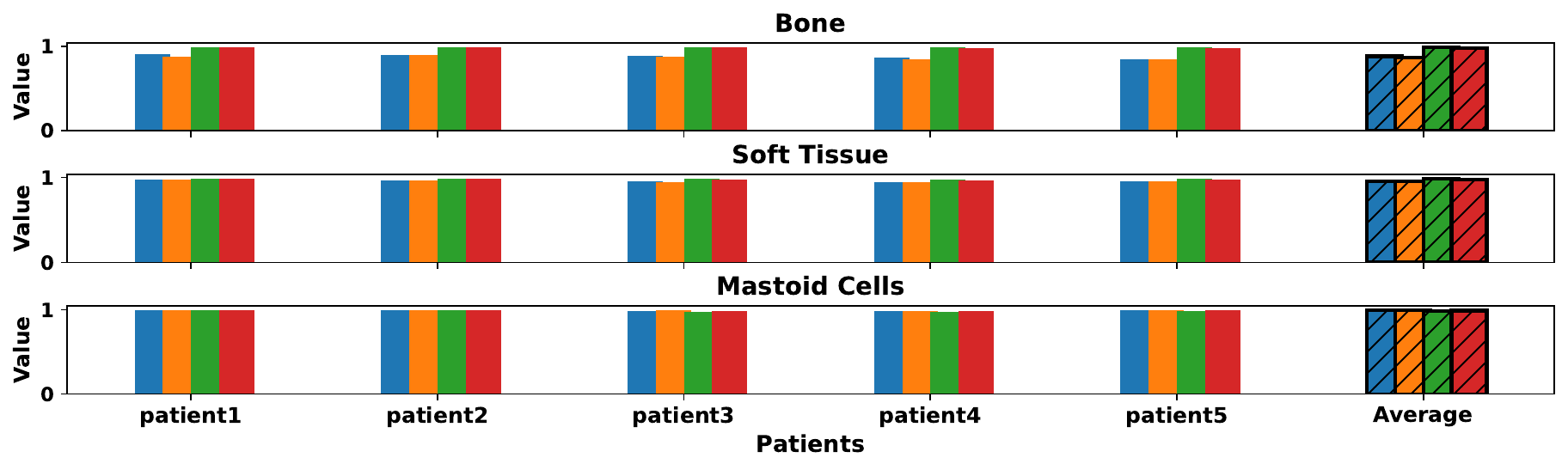} 
           \caption{Second approach}
           \label{fig:b}
       \end{subfigure}
       \hfill
       \begin{subfigure}{1.0\textwidth}
           \includegraphics[width=\textwidth]{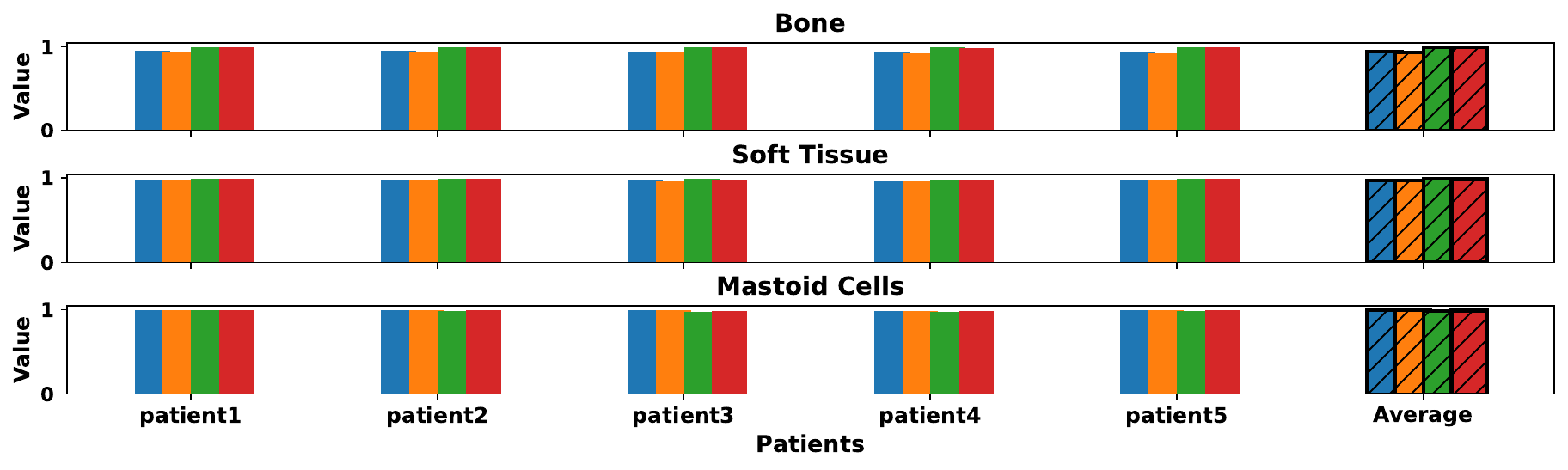} 
           \caption{Third approach}
           \label{fig:c}
       \end{subfigure}
       \caption{Dice coefficients, sensitivity, specificity, and accuracy for bone, soft tissue, and mastoid cells across five patient images, together with the average values over all patients, comparing the first, second, and third approaches with the Carney's method}
       \label{precision_figure}
\end{figure}

Table~\ref{precision_table} and Figure~\ref{precision_figure} summarize the Dice coefficients, sensitivity, specificity, and accuracy for bone, soft tissue, and mastoid cells, comparing the first, second, and third approaches with the Carney method. Mastoid cells demonstrated excellent agreement with the Carney AM, with Dice coefficients ranging from 0.99 to 1.00. Sensitivity was consistently high (0.99–1.00), indicating near-complete detection of mastoid air cells. Specificity ranged from 0.97 to 0.99, and overall accuracy was between 0.98 and 1.00 across all patient images and all three proposed approaches. Soft tissue segmentation also showed high agreement, with Dice coefficients ranging from 0.95 to 0.98 and differing within each patient only by 1\% for the three approaches. Sensitivity values varied between 0.95 and 0.98, while specificity remained consistently high at approximately 0.98–0.99 when compared with the three proposed approaches. Accuracy values were stable across patient images. Bone yielded slightly lower Dice coefficients, ranging from 0.85 to 0.95, with the second approach consistently producing the lowest values. Specificity was consistently high (0.99–1.00). Overall, the proposed approaches achieved high segmentation performance across all tissue classes and patient images. The highest agreement was observed for mastoid cells, followed by soft tissue, while bone segmentation showed slightly lower Dice scores but maintained high specificity and accuracy.

\begin{table}[ht!]
\caption{Dice coefficients, sensitivity, specificity, and accuracy for bone, soft tissue, and mastoid cells for the five patient images comparing the proposed approaches to Carney's method. A1, A2, and A3 are the first, second, and third approaches, respectively}
\label{precision_table}
\centering
\begin{tabular}{p{0.38cm}p{1.775cm}p{0.41cm} p{0.41cm}p{0.41cm}p{0.41cm}p{0.41cm}p{0.41cm}p{0.41cm}p{0.41cm}p{0.41cm}p{0.41cm}p{0.41cm}p{0.41cm}}
\toprule
\multirow{2}{0.38cm}{No.} &
\multirow{2}{1.775cm}{Tissue} &
\multicolumn{3}{c}{Dice Coefficient} &
\multicolumn{3}{c}{Sensitivity} &
\multicolumn{3}{c}{Specificity} &
\multicolumn{3}{c}{Accuracy} \\
\cmidrule(lr{.05cm}){3-5}
\cmidrule(lr{.05cm}){6-8}
\cmidrule(lr{.05cm}){9-11}
\cmidrule(lr{.05cm}){12-14}
& &
A1 & A2 & A3 &
A1 & A2 & A3 &
A1 & A2 & A3 &
A1 & A2 & A3 \\
\midrule
\text{P 1} 
& Bone           & 0.92 & 0.91 & 0.95 & 0.90 & 0.88 & 0.94 &  1.00 & 1.00 & 1.00 &  0.99 & 0.99 & 0.99\\
& Soft Tissue    & 0.97 & 0.97  & 0.98 & 0.98 & 0.98 & 0.98 & 0.99 &  0.98 & 0.99 & 0.98 & 0.98 & 0.99 \\
& Mastoid Cells  & 1.00 & 1.00 & 1.00 &  1.00 & 1.00 & 1.00 &  0.99 & 0.99 &0.99 & 0.99 & 0.99 &0.99 \\
\midrule
\text{P 2} 
& Bone           & 0.92 & 0.90 & 0.95 & 0.93 & 0.90 & 0.94 &  0.99 & 0.99 &1.00  & 0.99 & 0.99 & 0.99 \\
& Soft Tissue    & 0.97 & 0.96 & 0.98 &  0.96 & 0.96 & 0.98  & 0.99 & 0.99 & 0.99 & 0.98 & 0.98 & 0.99 \\
& Mastoid Cells  & 1.00 & 1.00  & 1.00 & 1.00 & 1.00 & 1.00 & 0.99 & 0.99 &0.99 & 1.00 & 1.00 & 1.00 \\
\midrule
\text{P 3} 
& Bone           & 0.91 & 0.89 & 0.94 & 0.92 & 0.88 & 0.93 & 0.99 & 0.99 & 1.00 & 0.99 & 0.99 & 0.99 \\
& Soft Tissue    & 0.96 & 0.95  & 0.97 & 0.95 & 0.95 & 0.96  & 0.99 & 0.98  &0.99 & 0.98 & 0.97 & 0.98 \\
& Mastoid Cells  & 0.99 & 0.99 & 0.99 &  0.99 &0.99  & 0.99 & 0.97 & 0.97 & 0.98 &  0.99 & 0.99 & 0.99\\
\midrule
\text{P 4} 
& Bone           & 0.87 & 0.86 & 0.93 & 0.87 & 0.85 & 0.92 & 0.99 & 0.99 &0.99 & 0.98 & 0.98 & 0.99 \\
& Soft Tissue    & 0.95 & 0.95 & 0.96 & 0.95 & 0.95 & 0.96 & 0.98 & 0.98  & 0.98 & 0.97 & 0.97 & 0.98 \\
& Mastoid Cells  & 0.99 & 0.99 & 0.99 & 0.99 & 0.99  & 0.99  & 0.98 & 0.98  & 0.97 & 0.98 & 0.98 &0.98 \\
\midrule
\text{P 5} 
& Bone           & 0.89 & 0.85 & 0.95 & 0.91 & 0.85 &0.93 & 0.99 & 0.99  & 1.00  & 0.99 & 0.98 & 0.99\\
& Soft Tissue    & 0.96 & 0.96 & 0.98 &  0.95 & 0.95 & 0.98& 0.99 & 0.99 &0.99 & 0.98 & 0.98 &0.99 \\
& Mastoid Cells  & 0.99 & 0.99 & 1.00  &  1.00 & 1.00  &1.00  & 0.99 & 0.99 & 0.99 & 0.99 & 0.99 & 0.99\\
\bottomrule
\end{tabular}
\end{table}

\subsubsection{Evaluation of differences}\label{differences in results}

\begin{figure}[ht!]
    \centering
    \includegraphics[width=1\textwidth]{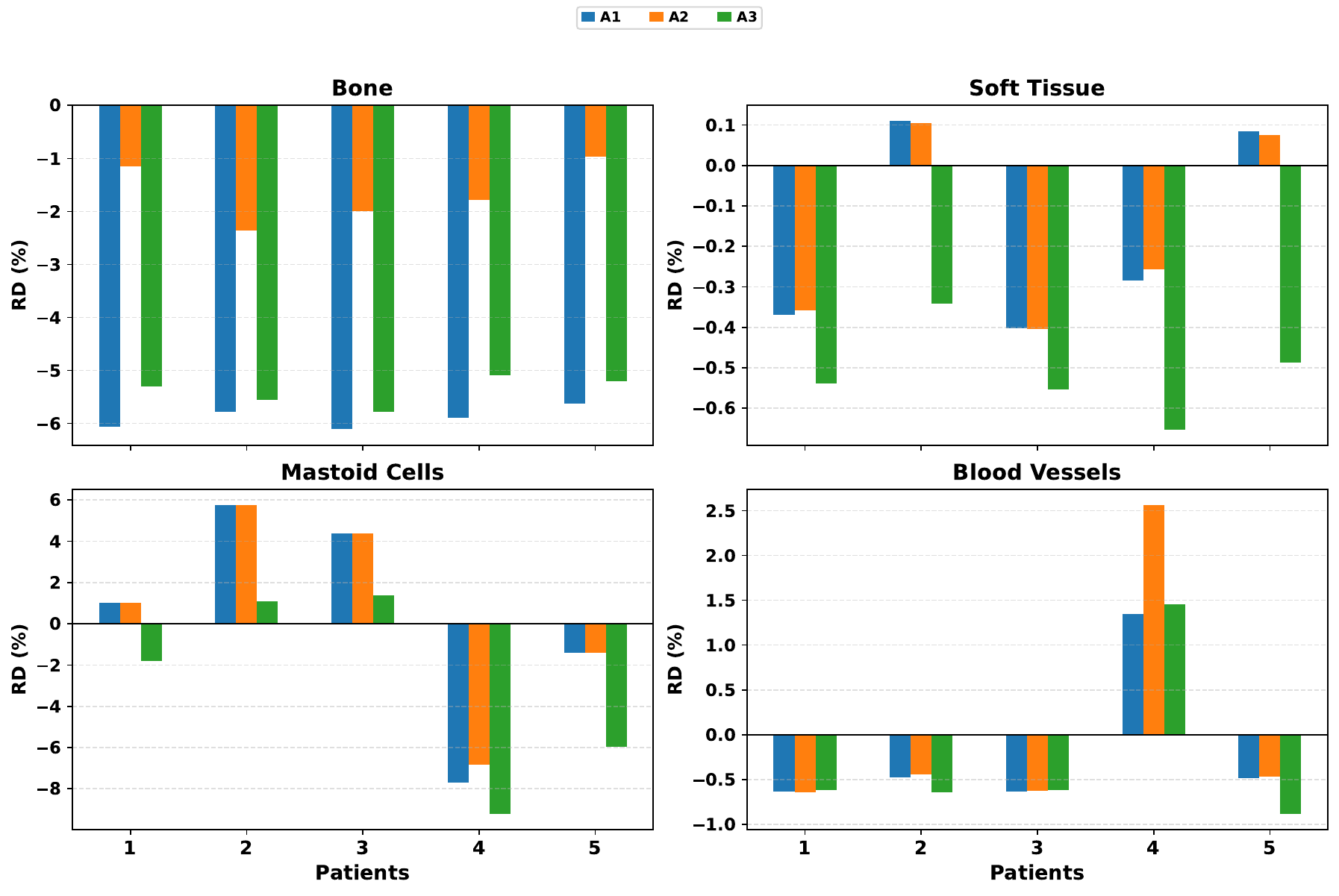}
    \caption{RD for bone, soft tissue, mastoid cells, and blood vessels for the five patient images comparing the proposed approaches to Carney's method. A1, A2, and A3 are the first, second, and third approaches, respectively}
    \label{RD_chart}
\end{figure}

Table~\ref{differences_table} presents the RD, RMSD, maximum, and minimum values for the first, second, and third approaches compared to the Carney method, evaluated across five patient images and four tissue classes. Figure~\ref{RD_chart} shows the RD values for the proposed approaches in the same patients. Among the tissue classes, bone shows the largest RDs for both the first and third approaches, while the second approach shows consistently lower RDs. In contrast to bone, mastoid cells present less systematic variability in RDs across patient images, including both positive and negative RDs, and, while low, slightly higher RMSD values than bone. 
In contrast, soft tissue and blood vessels exhibit very small RDs in most cases below 1\% and consistently low RMSD. The minimum and maximum values remain within a narrow range for all tissues and patient images. It is noticeable that the RDs for bone are negative across all the proposed approaches, indicating that the bones' LACs in our approaches are lower than those obtained using Carney's method. The RD values for soft tissue do not follow an apparent pattern. On the other hand, the RD values of blood vessels are negative for all patients except patient 4. Notably, patient 4 showed a low iodine concentration in the head and neck angiography images, which may explain this deviation.    

For the second approach, in comparison with the Carney method, soft tissue and blood vessels show comparable values to those observed in the first approach, which is expected since the same VNC images and conversion equation were used. A similar behaviour can be observed for the mastoid cells. It is also noticeable that the minimum values are identical when compared with the first approach.

For the third approach compared with the Carney method, bone shows comparable values to those of the first approach, which is reasonable since the same VNI images and equation were applied.

\clearpage

\begin{sidewaystable}
\centering
\caption{RD, RMSD, minimum, and maximum values for bone, soft tissue, mastoid cells, and blood vessels for the five patient images comparing the proposed approaches to Carney's method. A1, A2, and A3 are the first, second, and third approaches, respectively}
\label{differences_table}
\begin{tabular*}{0.95\textheight}{@{\extracolsep\fill}p{0.38cm}p{1.9cm}cccc cccc cccc cccc}
\toprule
\makecell{No} &
\makecell{Tissue} &
\multicolumn{3}{c}{RD (\%)} &
\multicolumn{3}{c}{RMSD} &
\multicolumn{3}{c}{Maximum} &
\multicolumn{3}{c}{Minimum} \\
\cmidrule(lr){3-5}
\cmidrule(lr){6-8}
\cmidrule(lr){9-11}
\cmidrule(lr){12-14}
& &
A1 & A2 & A3 &
A1 & A2 & A3 &
A1 & A2 & A3 &
A1 & A2 & A3 \\
\midrule
\text{P 1} 
& Bone           & -6.066 &  -1.156  & -5.303 & 0.013 &  0.015  & 0.012 & 0.065 &  0.086  & 0.051 & -0.124 &  -0.124  & -0.101 \\
& Soft Tissue    & -0.369 &  -0.359  & -0.538 & 0.004 &  0.004  & 0.003 & 0.081 &  0.103   & 0.050 & -0.106 & -0.106   & -0.092 \\
& Mastoid Cells  & 1.020 &  1.035  & -1.801 & 0.019 & 0.019   & 0.013 & 0.119 &  0.123  & 0.058 & -0.086 &  -0.086  & -0.081 \\
& Blood Vessels  & -0.635 & -0.642 &   -0.618 &  0.003 &  0.003  & 0.003 & 0.062 &  0.064   & 0.027  & -0.080 &  -0.080  & -0.056 \\
\midrule
\text{P 2} 
& Bone           & -5.786 &  -2.357  & -5.556 & 0.011 &   0.012 & 0.011 & 0.041 & 0.070   & 0.037 & -0.104 &  -0.104  & -0.100\\
& Soft Tissue    & 0.110 &  0.104  & -0.342 & 0.004 &  0.004  & 0.003 & 0.037 & 0.072 & 0.037 & -0.099 &  -0.099  & -0.094 \\
& Mastoid Cells  & 5.761 & 5.759   & 1.101 & 0.017 &  0.017  & 0.013 & 0.071 & 0.073   & 0.051 & -0.086 &  -0.086  & -0.079\\
& Blood Vessels  & -0.477 &  -0.442  & -0.646 & 0.004 &  0.004  & 0.003  & 0.040 & 0.040   & 0.041  & -0.079 &  -0.079  & -0.078 \\
\midrule
\text{P 3} 
& Bone           & -6.110 &  -1.995  & -5.777 & 0.013 &  0.014  & 0.013 & 0.068 & 0.085   & 0.068  & -0.113&  -0.113  & -0.106  \\
& Soft Tissue    & -0.403 & -0.405   & -0.554 & 0.005 &  0.005  & 0.004 & 0.070 &  0.092  & 0.070  & -0.103&  -0.103  & -0.102  \\
& Mastoid Cells  & 4.393 & 4.395   & 1.388 & 0.022 &  0.022  & 0.019 & 0.082 &  0.098   & 0.052 & -0.089&  -0.089  & -0.089  \\
& Blood Vessels  & -0.633 &  -0.624  & -0.620 & 0.004 & 0.004   & 0.003 & 0.046 &  0.046  & 0.042 & -0.075 &  -0.075  & -0.059\\
\midrule
\text{P 4} 
& Bone           & -5.897 & -1.778   & -5.088 & 0.014 &  0.016  & 0.013 & 0.051 &  0.081  & 0.051 & -0.144 &  -0.144  & -0.125\\
& Soft Tissue    & -0.285 &  -0.256  & -0.654 & 0.006 &  0.006  & 0.006 & 0.061 &  0.076  & 0.059 & -0.109 &  -0.109  &  -0.113\\
& Mastoid Cells  & -7.719 &  -6.820  & -9.233 & 0.032 &  0.033   & 0.031  & 0.100 &  0.106  & 0.085 & -0.089 & -0.089   & -0.097\\
& Blood Vessels  & 1.347 &  2.564  & 1.456 & 0.008 &  0.012  & 0.008 & 0.093 & 0.106   & 0.072 & -0.065 &  -0.065  & -0.055 \\
\midrule
\text{P 5} 
& Bone           & -5.627 &  -0.973  & -5.197 & 0.011 & 0.015   & 0.010 & 0.056 &  0.073  & 0.036 & -0.116 & -0.116   & -0.080\\
& Soft Tissue    & 0.084 &  0.075  & -0.487 & 0.004 &  0.004   & 0.003 & 0.050 &  0.077  & 0.036 & -0.106 & -0.106   & -0.085 \\
& Mastoid Cells  & -1.417 &  -1.420  & -5.958  & 0.022 & 0.022   & 0.017 & 0.079 &  0.089   & 0.051 & -0.091 & -0.091   & -0.077 \\
& Blood Vessels  & -0.486 &  -0.470   & -0.888  & 0.005 & 0.005    & 0.004 & 0.031 &  0.031  & 0.023 & -0.083 &  -0.083  & -0.061 \\
\bottomrule
\end{tabular*}

\end{sidewaystable}

\clearpage

\section{Discussion}\label{Discussion}

In this study, a transformation scheme for converting measured HUs to LACs for gamma photons at \SI{511}{\keV} was developed by acquiring spectral CT data for different tissue types using the Sun Nuclear electron density phantom. The transformation scheme was designed to create an attenuation template from angiographic CT images with iodinated contrast agent for a custom-made, human, 7T MR-compatible BrainPET insert \cite{lercheDesignPerformanceEstimation, lercheFirstPerformanceResults2023}.

Three approaches for obtaining transformation schemes from iodine-enhanced spectral CTs were proposed. In the first approach, VNC images were combined with the bone from the VNI images; the second relied only on VNC images; and the third used only VNI images. Similar to the kVp dependence described by Carney et al.~\cite{carneyMethodTransformingCT2006}, there is a significant dependency on the virtual CT energy for the VNI and the combined VNI/VNC approaches. This energy dependence appears to replace the kVp dependence when using virtual monoenergetic images (see Table~\ref{VNI_table}). For these cases, energy-dependent, but not tube voltage-dependent, transformation schemes are required. As in Carney's case, for HU values close to and below that of water, all three approaches exhibit a nearly identical, energy-independent behaviour.

The three approaches were further evaluated using five patient images and compared to Carney's method. For this evaluation, accuracy, sensitivity, specificity, Dice coefficients, RDs, RMSD, maximum, and minimum values were computed for bone, soft tissue, and mastoid cells. The AMs from all three approaches demonstrated good agreement when compared to the Carney method using native head CT images. Consequently, unaltered quantitation accuracy can be expected for PET images acquired using this template in a PET/MR machine. While all three approaches show similar accuracy, the second approach appears to perform slightly better. In all three approaches, the iodinated contrast agent is effectively removed from the blood. Variability in the mastoid cells across patient images likely reflects their complex anatomy and segmentation uncertainties. The presence of both positive and negative RDs indicates that these deviations are patient-specific rather than systematic. Notably, the mastoid cells in patient~4 exhibit a higher RD across all approaches. The results demonstrate that the the overall fraction of voxels corresponding to blood vessels is rather small (see blood vessels for patient 5 in Table~\ref{differences_table}), and a bias is expected, as discussed in \cite{hamillTechnicalNoteCan2024}. Similarly, a straightforward application of Carney's method to the VMI images is likely to introduce bias. 


Interestingly, the second approach, which relies on the VNC images and the corresponding equations, yields the lowest RD, even though calcium is partially subtracted from the VNC images. As a result, the HU values of the bone in the VNC images may be less accurate than those in the VNI image. This necessarily leads to very different HU values for bone in VNC images when compared to conventional CT images. However, provided consistent CT acquisition and image reconstruction settings are used when deriving both the transformation and the attenuation template, the accuracy of the AM will not be impacted. The VNC approach is also of special interest for AC, as it provides a single linear function across the entire HU range that is independent of the tube voltage. In this context, it is remarkable, that the data points of the soft tissue compartment follow linear mapping more closely than with Carney's method (the error bars in Figure~\ref{Extrapolation schemes} represent the standard error of the mean for the HU values were multiplied by 600 to be visible), where soft tissue residuals between linear transformation values and the discrete HU and LAC values are much larger. Similar observations can also be made for the first and third approaches. Although this feature should be further evaluated as a source of ground truth attenuation values (i.e.~with transmission measurements using a PET-only scanner), it may offer a more precise method for CT-based AC in PET/CT if the CT component offers DECT and VNC images. This is especially relevant for CT images without an iodinated contrast agent, as the VNC approach is expected to work equally well with and without iodinated blood. The latest generation of scanners, such as the Biograph Vision Quadra PET/CT scanner support this capability \cite{VisionQuadra}. In contrast, only using VMI images does not lead to the same effect, as reported by Hamill et al.~\cite{hamillTechnicalNoteCan2024} for VMI images computed at very high energies.
While Hamill reported that the LACs derived from PCCT-based measurements agreed within \SI{3}{\percent} for inserts containing iodine, the proposed VNC-based transformation scheme demonstrated significantly smaller relative residuals (below \SI{1}{\percent}) for most materials, except lung tissue. This is likely due to the difficulty in accurately mimicking lung tissue, given its inherently non-uniform mechanical properties \cite{Caro_Pedley_Schroter_Seed_Parker_2011} and the non-uniform mechanical structure of the lung inserts, i.e.~gaseous inclusions \cite{OurPreviousStudy}. Additionally, the larger residuals were observed for air (see Table~\ref{relative_residuals_VNC_120kV}), arising from division by values close to zero; however, the residuals' difference remains small. Therefore, our approach, particularly the VNC scheme, shows improved agreement and performance accuracy compared to the results reported by Hamill et al.~\cite{hamillTechnicalNoteCan2024}. Furthermore, we found that the transformation scheme proposed by Carney leads to an overestimation of the linear attenuation for the low density bone reference tissues of up to \SI{6}{\percent} (see the discussion section in \cite{carneyMethodTransformingCT2006}).

This study has several limitations. First and foremost we cannot compare with the ground truth, since it is not available. Instead, we made a comparison with the state of the art (i.e.~the Carney method), which also has deficiencies. Additionally, the AMs were not tested in emission reconstructions, since the template and system are still under development.

\section{Conclusion}\label{Conclusion}

In this work, we investigated the feasibility of using contrast-enhanced spectral CT angiography data for attenuation map generation at \SI{511}{\keV}. The results demonstrate that angiographic datasets acquired with iodinated contrast can be used for AC template generation for PET/MR imaging and potentially also for PET/CT imaging. Three approaches were proposed to obtain the attenuation maps. Overall, good agreement with the Carney method was observed, especially for soft tissue and blood vessels, which showed minimal relative differences and low RMSD values across all three proposed approaches. In contrast, bone exhibited larger deviations. 

\bibliography{sn-bibliography}

\bmhead*{Acknowledgements} These data are part of the Dr.~rer.~medic.~thesis at the Medical Faculty of the RWTH Aachen University, Germany, of Heba R.~S.~Alrakh. The authors would like to thank Claire Rick for proofreading the manuscript.

\section*{Statements and Declarations} 

\bmhead*{Funding} This work is funded by the Palestinian-German Science Bridge (PGSB), the RETIMAGER project (Real Time Molecular Imager with Unsurpassed Resolution), and the Innovation Go fund by Forschungszentrum Jülich.

\bmhead*{Competing interests} Sebastian Faby is employed by Siemens Healthineers AG and has no control over the data. The other authors have no relevant financial or non-financial interests to disclose.

\bmhead*{Ethics approval and consent to participate} Not applicable.

\bmhead*{Consent for publication} Not applicable.

\bmhead*{Data availability} The datasets generated during and/or analysed during the current study are available from the corresponding author on reasonable request.

\bmhead*{Ethics approval} The study was reviewed and approved by EK 24-270 (Ethikkommittee der Medizinischen Fakultät, Universität RWTH Aachen).

\bmhead*{Materials availability} The patients' images will not be available to protect their privacy. 

\bmhead*{Code availability} The code of the current study is available from the corresponding author on reasonable request.

\bmhead*{Author contribution} Heba Alrakh: conceptualization, literature review, material preparation, data collection, analysis, writing original draft, visualization, review, and editing. Lutz Tellmann: material preparation and data collection. Omid Nikoubashman: assisting in selecting the appropriate parameters for the photon-counting CT (PCCT). Sadaf Shokrgozar: providing support in performing the experimental measurements. Sebastian Faby: contribution to the development of the methodology. Martin Wiesmann: providing valuable guidance on spectral CT. Cláudia Régio Brambilla, Ahlam Said Mohamad Issa, Keith George Ciantar, Ali Abu arra, and N.~Jon Shah: review. Christoph Lerche: literature review, material preparation, data collection, review, supervision, project supervision, and funding. All authors contributed to the final approval of the manuscript and agree to be accountable for its content.

\end{document}